\documentclass[showpacs, twocolumn]{revtex4-1}
\usepackage{graphicx}
\usepackage{amsmath}

\begin{document}

\title{Self-localized state and solitons in a Bose-Einstein-condensate-impurity mixture at finite temperature}

\author{Abdel\^{a}ali Boudjem\^{a}a}

\affiliation{Department of Physics, Faculty of Sciences, Hassiba Benbouali University of Chlef P.O. Box 151, 02000, Ouled Fares, Chlef, Algeria.}

\email {a.boudjemaa@univ-chlef.dz}

%\date{\today}

\begin{abstract}
We study the properties of a Bose-Einstein condensate (BEC)-impurity mixture at finite temperature
employing the time dependent Hartree-Fock Bogoliubov (TDHFB) theory which is a set of coupled nonlinear equations of motion for the condensate and its 
normal and anomalous fluctuations on the one hand, and for impurity on the other. 
The numerical solutions of these equations in the static quasi-1D regime show that the thermal cloud and the anomalous density are deformed as happens to the condensate
and the impurity becomes less localized at nonzero temperatures.
Effects of the BEC fluctuations on the self-trapping state are studied in  homogeneous weakly interacting BEC-impurity at low temperature. 
The self-trapping threshold is also determined in such a system.
The formation of solitons in the BEC-impurity mixture at finite temperature is investigated. Our formalism shows several new pictures.
\end{abstract}


\maketitle

\section{Introduction} \label{Intro}
During recent years, a revived interest in BEC-impurity mixtures has been stimulated by the experimental works of the authors of Refs.\cite {Chik, Ciam, Gunter, Ospe}.
In particular, it has been proven that single atoms can get trapped in the localized distortion of the BEC that is induced by the impurity-BEC interaction \cite{Pal, Brud, Will}.
Recently, Catani et\textit {al}.\cite {Cat} created a harmonically trapped impurity suspended in a separately trapped Bose gas 
and they studied the dynamics of such a system following a sudden lowering of the trap frequency of the impurity.
Very recently, an important exprerimental study of the quantum dynamics of a deterministically created spin-impurity atom propagated 
in a one-dimensional lattice system has been realized in Ref. \cite{bloch}.  

Theoretically, the self-trapping impurities in BEC with strong attractive and repulsive coupling have been studied 
in  homogeneous and harmonically trapped condensate \cite {Jack1, Tim1}.
The quasiparticle excitation spectrum  and quantum fluctuations around the product state that describes the entanglement of the impurity and boson
degrees of freedom, have been calculated in a homogeneous case \cite{Tim2}. In such a system,  the formation of a parametric soliton behavior has also been predicted \cite {Tim1}.  
Moreover, it has been shown that the self-localized BEC-impurity state resembles that of a small polaron which  has been described successfully in the strong coupling 
limit using both the Landau-Pekar treatment \cite {Tim3} and the Fr\"o\-hlich-Bogoliubov Hamiltonian within the Feynman path-integral
\cite {Tim4,Temp}. Then, this study was generalized to two polaron flavors and multi-impurity polarons in a dilute BEC by Tempere \textit {al}.\cite {Tim4} and Blinova et \textit {al}. \cite {Tim5}. 
Furthermore, the dynamics and the breathing oscillations of a trapped impurity  as well as the impurity transport through a strongly interacting 
bosonic quantum gas are investigated in Refs.\cite {Jack2, Jack3}.
Additionally, the properties of the impurity-BEC in a double well potential are discussed in \cite {Dell}. 

Although these theories give good results at zero temperature, they completely ignore  the behavior of BEC-impurity at finite temeprature. 
The effects of the temperature are so important, in particular on the fluctuations, on the expansion of the condensate, and on the thermodynamics of the system.
Certainly, the dynamics of the BEC-impurity  at nonzero temperatures is a challenging problem as for example the Bogoliubov approximation becomes invalid, at least at large times, and large
thermal phase fluctuations have to be taken into account even at low temperatures where density fluctuations are small.
It is therefore instructive to derive a self consistent approach to describe the static and the dynamic behavior of BEC-impurity mixtures
at finite temperature  especially because all experiments actually take place at nonzero temperatures.

Our approach is based on the time-dependent Balian and M. V\'en\'eroni  (BV)  variational principle \cite {BV}. 
This variational principle requires first the choice of a trial density operator.
In our case, we consider a Gaussian time-dependent density operator. This ansatz belongs to the class
of the generalized coherent states. The BV variational principle is based on the minimization of an action 
which involves two variational objects : one is related to the observables of interest
and the other is akin to a density matrix \cite {Ben}. This leads to a set of coupled time-dependent mean-field equations 
for the condensate, the noncondensate, the anomalous average and the impurity. This approach is called “time-dependent Hartree-Fock-Bogoliubov” (TDHFB). 

The original numerical implementation of this theory \cite {boudj2010} successfully addressed the issue of the condensate and the thermal cloud formation at finite temperature. 
Likewise, the TDHFB equations have been used  to study the properties of the so-called anomalous density in three and two-dimensional homogeneous and trapped Bose gases \cite {boudj2011, boudj2012}.  The results of this analysis present an overall good agreement with recent experimental and theoretical works and highly coincide with the Monte Carlo simulation. 
The TDHFB theory yields also remarkable agreement with various experiments, e.g., hydrodynamic collective modes and vortex nucleation at finite temperature \cite {boudj 2013}. 

The rest of this paper is organized as follows. In Sec.\ref {Model}, we review the main steps used to derive the TDHFB equations from
the BV variational principle. In Sec. \ref {appl}, the TDHFB equations are applied to a trapped BEC-impurity system to derive
a set of coupled equations governing the dynamics of the condensate, the noncondensate, the anomalous average, and the impurity. We then restrict ourselves to 
solve these equations numerically in a static quasi-1D case and we therefore, look at how much the impurity enhances the condensate fluctuations 
and how much it may be localized.
In Sec.\ref{BECflu}, we discuss the effects of the condensate fluctuations on the self-trapping impurity using the linearized TDHFB equations in a homogeneous quasi-1D case. 
Formulas of some thermodynamic quantities of such a system are also given. 
Section.\ref{sol} is devoted to studying the behavior and the formation of solitons in BEC-impurity mixtures in quasi-1D geometry.
In this section we analyze numerically  the different scenarios that emerge in our model, as well as the temperature effects on the depth and on the creation of solitons.
Finally we present our conclusions in Sec.\ref {conc}.

\section {TDHFB equations} \label{Model}
The Gaussian density operator ${\cal D}(t)$ is completely
characterized by the partition function ${\cal Z}(t)=\hbox{Tr}\,{\cal D}(t)$, 
the one boson field expectation value $\langle \hat\psi \rangle ({\bf r}, t)=\hbox{Tr}\,\hat\psi({\bf r})\, {\cal D}(t) /{\cal Z}(t) $ and the single particle density matrix is defined as 
\begin{equation}
\rho_j ({\bf r},{\bf r'},t)=\begin{pmatrix} 
\langle \hat{\bar{\psi}}^{+}\hat{\bar{\psi}}\rangle & -\langle\hat{\bar{\psi}}\hat{\bar{\psi}}\rangle\\
\langle\hat{\bar{\psi}}^{+}\hat{\bar{\psi}}^{+}\rangle& -\langle\hat{\bar{\psi}}\hat{\bar{\psi}}^{+}\rangle
\end{pmatrix}_j ({\bf r},{\bf r'},t)
\end{equation}
where $j$ refers to the BEC atoms as ‘B’ and to the impurity neutral atoms as ‘I’.\\
In the preceding definitions, $\hat\psi_j $ and $\hat\psi_j^{+} $ are the boson destruction and creation field operators (in the Schr\"o\-dinger
representation), respectively, satisfying the usual canonical commutation rules 
$[\hat\psi_j({\bf r}), \hat\psi_j^{+}(\bf r')]=\delta ({\bf r}-{\bf r'})$ and $\hat{\bar \psi}_j({\bf r})=\hat\psi_j({\bf r})- \Phi_j({\bf r})$ is the noncondensed part of the field operator with $\Phi_j=\langle\hat\psi_j({\bf r})\rangle$.

Upon introducing these variational parameters into the BV
principle, one obtains dynamical equations for the expectation
values of the one- and two-boson field operators \cite {Ben, boudj2010, boudj2011}
\begin{equation} \label {eq1}
i\hbar \frac{d  \Phi_j}{d t} =\frac{d{\cal E}}{d \Phi_j},
\end{equation}
\begin{equation} \label {eq2}
i\hbar \frac{d \rho_j}{d t} =\left[\rho_j, \frac{d{\cal E}}{d\rho_j^{+}}\right].
\end{equation}
One of the most noticeable properties of these equations is the unitary evolution of the one-body density matrix $\rho_j$, which means
that the eigenvalues of $\rho_j$ are conserved. This immediately leads to the expression 
\begin{equation} \label{Invar}
\rho_j (\rho_j +1)= ( I_j-1)/4.
\end{equation} 
where $I$ known as the Heisenberg invariant.\\
Therefore, Eq.(\ref{Invar}) involves the conservation of the von Neumann entropy $S = −\hbox{Tr}\,{\cal D} \ln{\cal D}$. \\
Indeed, parameter (\ref{Invar}) is related to the degree of mixing (see Appendix A of Ref. \cite{Cic}). For pure state  and at zero temperature, $I=1$.

Among the advantages of the TDHFB equations is that they should yield the general time,
space, and temperature dependence of the various densities.
Furthermore, they satisfy the energy and number conserving laws.  Interestingly, our TDHFB
equations can be extended to provide self-consistent equations of motion for the triplet correlation function by using the
post-Gaussian ansatz.

\section{Application to the BEC-impurity system} \label{appl}
We consider $N_I$ impurity atoms of mass $m_I$ in an external trap $V_I({\bf r})$, and identical bosons of mass $m_B$ trapped by an external potential $V_B({\bf r})$. The impurity-boson interaction and boson-boson interactions have been approximated by the contact potentials $g_B \delta ({\bf r}-{\bf r'})$ and $g_{IB} \delta ({\bf r}-{\bf r'})$, respectively. We neglect the mutual interactions of impurity atoms under the assumption that their number and local density remains sufficiently small \cite {Tim1, Jack1}. 
The many-body Hamiltonian for the combined system which  describes  bosons, impurity and impurity-boson gas coupling is given by 
\begin{eqnarray} \label{eq4}
&& \hat H =\hat H_B+\hat H_I+\hat H_{IB} \\&& \nonumber
=\int d{\bf r} \hat\psi_B^{+}({\bf r}) \left[-{\frac{\hbar^2}{ 2m_B}}\Delta + V_B({\bf r}) +\frac{g_B}{2} \hat \psi_B^{+}({\bf r})\hat \psi_B({\bf r})\right]\hat\psi_B({\bf r}) \\&& \nonumber
+\int d{\bf r} \hat\psi_I^{+}({\bf r}) \left[-{\frac{\hbar^2}{ 2m_I}}\Delta + V_I({\bf r})\right]\hat\psi_I({\bf r})\\&& \nonumber
+g_{IB}\int d{\bf r} \hat\psi_I^{+}({\bf r}) \hat \psi_I({\bf r})\hat \psi_B^{+} ({\bf r}) \hat\psi_B({\bf r}),
\end{eqnarray}
where  $\hat\psi_B(\bf r)$ and $\hat\psi_I({\bf r})$ are the boson and impurity field operators.

The total energy ${\cal E}={\cal E_B}+{\cal E_I}+{\cal E_{IB}}=\langle \hat H\rangle$ can be easily computed yielding: 
\begin{subequations} \label{E:eng}
\begin{align} 
&  {\cal E_B} = \int d{\bf r} \left (-{\frac{\hbar^2}{2m_B}}\Delta + V_B \right) (|\Phi_B|^2+\tilde{n}) \nonumber \\ 
&+\frac{g_B}{2} \int d{\bf r} \left(|\Phi_B|^4+ 4\tilde{n}|\Phi_B|^2+2\tilde{n}^2 +|\tilde{m}|^2 + \tilde{m}^{*}\Phi_B^2 + \tilde{m} {\Phi_B^{*}}^2\right) , \\  
&{\cal E_I} = \int d{\bf r}\left[\left ( -{\frac{\hbar^2}{2m_I}}\Delta + V_I \right) (|\Phi_I|^2+\tilde{n}_I)\right], \\ 
&{\cal E_{IB}} = g_{IB}\int d{\bf r} (|\Phi_I|^2+\tilde{n}_I) (|\Phi_B|^2+\tilde{n}),
\end{align}
\end{subequations}
where $\Phi_B$ and $\Phi_I$ stand for the condensate and the impurity wave functions, respectively.  The noncondensed density $\tilde{n}$ and the anomalous density $\tilde{m}$
are identified, respectively, with $\langle\hat {\bar {\psi}}_B^{+}\hat{\bar {\psi}}_B\rangle$, $\langle\hat {\bar {\psi}}_B\hat{\bar {\psi}}_B\rangle$ 
and $\tilde{n}_I=\langle\hat{\bar {\psi}}_I^{+}\hat{\bar {\psi}}_I\rangle$ is the impurity fluctuation. 

Expressions (\ref{E:eng}) for the energy allow us to write down Eqs.(\ref {eq1}) and (\ref {eq2}) more explicitly as \\
\begin {widetext} 
\begin{subequations}\label{E:gp}
\begin{align} 
& i\hbar \dot{\Phi}_B  = \left( -{\frac{\displaystyle\hbar^2}{\displaystyle 2m_B}}\Delta + V_B +g_B (|\Phi_B|^2+2\tilde{n}) + g_{IB} (|\Phi_I|^2+\tilde{n}_I)  \right)\Phi_B + g_B\tilde{m}
\Phi_B^{*} ,  \label{E:gp1} \\  
&i\hbar \dot{\Phi}_I  = \left( -{\frac{\displaystyle\hbar^2}{\displaystyle 2m_I}}\Delta + V_I +g_{IB} (|\Phi_B|^2+\tilde{n})\right)\Phi_I ,  \label{E:gp2} \\ 
&i\hbar \dot{\tilde{n}}  = g_B\left(\tilde{m}^{*}\Phi_B^2-\tilde{m} {\Phi_B^{*}}^2\right) ,  \label{E:gp3} \\ 
&i\hbar \dot{\tilde{n}}_I  = 0 ,   \label{E:gp4}\\ 
&i\hbar \dot{\tilde{m}} = g_B (2\tilde{n} +1)\Phi_B^2 + 4\left( -{\frac{\displaystyle\hbar^2}{\displaystyle 2m_B}}\Delta + V_B+2g_B n+{
\frac{\displaystyle g_B}{\displaystyle 4}}(2\tilde{n} +1)+g_{IB} (|\Phi_I|^2+\tilde{n}_I) \right)\tilde{m}   \label{E:gp5},
\end{align}
\end{subequations}
\end {widetext}
where $n=|\Phi_B|^2+\tilde {n}$ is the total density in the BEC.\\ 
Putting $g_{IB} =0$ (i.e., neglecting the mean-field interaction energy between bosons and impurity components)
one recovers the usual TDHFB equations \cite {boudj2010, boudj2011, boudj2012, boudj 2013} describing a degenerate Bose gas at finite temperature
and the Schr\"o\-dinger equations describing a noninteracting impurity system. 
In the case when $\tilde{n}=\tilde{m}=0$, Eqs (\ref {E:gp}) becomes similar to those derived in Ref. \cite{karp} for Bose-Fermi mixtures
with fermions playing the role of the impurity.\\
Interestingly, we see from Eq.(\ref {E:gp4}) that the noncondensed density of the impurity is constant while
the anomalous density which describes correlations between pairs does not exist in such a system.  Indeed, the absence of the anomalous density in the impurity is due to the neglect of the interaction between impurity atoms. One should mention also at this level that Eq.(\ref {E:gp5}) which describes the behavior of the anomalous density-impurity has no analog in the literature.
%and, as known, the pairing strongly depends to the interactions \cite{boudj2011, Burnet}.

A useful link between the noncondensed and the anomalous densities of BEC can be given via Eq.(\ref {Invar}) 
\begin{align} \label{eq6}
I_B& =(2\tilde{n} +1)^2-4|\tilde{m} |^2.  
%I_I &=(2\tilde{n}_I +1)^2.
\end{align}
Equation (\ref{eq6}) clearly shows that $\tilde{m}$ is larger than $\tilde{n}$ at low temperature, so the omission of the anomalous density in this situation 
is principally an unjustified approximation and wrong from the mathematical point of view.\\
Notice that for a thermal distribution, $I_k=\coth^2 ( \varepsilon_k/T)$,  where $\varepsilon_k$ is the excitation energy of the system. 
The expression of $I$ allows us to calculate in a very useful way the dissipated heat for the \textit{d}-dimensional BEC-impurity mixture as $Q=(1/n)\int E_k I_k d^dk/(2\pi)^d$ with $ E_k=\hbar ^2k^2/2m$ \cite {boudj2012}. It is necessary to stress also that our formalism provides an interesting formula for the superfluid fraction $n_s=1-2Q/d T$ \cite {boudj2012} which reflects the importance of the parameter $I$.

Equations (\ref {E:gp}) in principle cannot be used as they stand since they do not guarantee to give the best excitation frequencies. Indeed it is well know \cite {Gir, Burnet, boudj2011} that the inclusion of the anomalous average leads to a theory with a (unphysical) gap in the excitation spectrum. The standard treatment in calculations for trapped gases has been to neglect $\tilde {m}$ in the above equations, which restores the symmetry and hence leads to a gapless theory. This is often reminiscent of the Popov approximation.
In addition, one finds that the anomalous average is divergent if one uses a contact interaction. To go beyond Popov, one has to renormalize the anomalous average to  circumvent
this ultraviolet divergence. Following the method described in Ref. \cite{Burnet}, we get the following from Eq.(\ref {E:gp1})
\begin{align} \label{CC}
&g_B |\Phi_B|^2\Phi_B+g_B\tilde{m}\Phi_B^{*}=g_B(1+\tilde {m}/\Phi_B ^2)|\Phi_B|^2\Phi_B 
\\=& U |\Phi_B|^2\Phi_B \nonumber.
\end{align} 
This is similar to the so-called $G2$ approximation based on the $T$-matrix calculation  \cite{Burnet}.  \\ 
At very low temperature where $\tilde {m}/\Phi_B ^2 \ll 1$, the new coupling constant $U$ reduces immediately to $g_B$. \\
Inserting $U$ in Eqs. (\ref {E:gp}) and using $2\tilde {n}+1\approx 2\tilde {m}$ \cite{boudj 2013}, this approximation is valid only at very low temperature
where $\tilde{m} \geq \tilde{n}$ as we have mentioned above. After some algebra we obtain
\begin {widetext} 
\begin{subequations}\label{E:td}
\begin{align}
&i\hbar \dot{\Phi}_B  = \left \{-{\frac{\displaystyle\hbar^2}{\displaystyle 2m_B}}\Delta + V_B +g_B \left[\beta |\Phi_B|^2+2\tilde{n}+\gamma(|\Phi_I|^2+\tilde{n}_I)\right] \right\}\Phi_B,  \label{E:td1}
\\ 
&i\hbar \dot{\tilde{m}} = \left\{ -{\frac{\displaystyle\hbar^2}{\displaystyle 2m_B}}\Delta + V_B+g_B \left[2G \tilde {m}+2n+\gamma(|\Phi_I|^2+\tilde{n}_I)\right]\right\}\tilde{m}, \label{E:td2}
\end{align}
\end{subequations}
\end {widetext}
where $\beta = U /g_B$, $G = \beta / 4(\beta-1 )$ and $\gamma=g_{IB}/g_B$.\\
Let us now reveal the significance  of parameter $\beta$. First of all, $\beta$ accounts for finite-temperature effects (dissipation), it scales with temperature $T$ according to the formula (\ref{eq6}).
Futhermore, for $\beta = 1$, i.e., $ \tilde{m} /\Phi_B^2 = 0$, Eq.(\ref {E:td1}) reduces to the well-known HFB-Popov
equation which is safe from all ultraviolet and infrared divergences and thus provides a gapless spectrum. 
For $0<\beta<1$, $G$ is negative %and hence, $\tilde{m} $ has a negative sign. 
while for $\beta >1$, $G$ is positive. %, and thus, $\tilde{m} $ becomes a positive quantity.
%Indeed,  the significance of the anomalous density is ascribed to its modulus but not to its sign.
At this level, we note that for large values of $\beta$, one gets a BEC with strong interactions and high correlations. 
So,  in order to guarantee the diluteness of the system, $\beta$ should vary as $\beta=1\pm\epsilon$ with $\epsilon$ being a small value.

In what follows we consider a single impurity $N_I=1$, which means that there is no impurity fluctuation ($\tilde{n}_I=0$), immersed in elongated (along the $x$-direction) BEC and confined in a highly anisotropic trap (such that the longitudinal and transverse trapping frequencies are $\omega_{Bx}/\omega_{B\perp}\ll1$). 
In such a case, the system can be considered as quasi-1D and, hence, the coupling constants of the Hamiltonian (\ref{eq4}) effectively
take their 1D form, namely $g_B =2\hbar\omega_{B\perp} a_B$ and $g_{IB} =2\hbar\omega_{B\perp} a_{IB}$, where  $a_B$ and $a_{IB}$ are the scattering lengths describing
the low energy boson-boson  and impurity-boson scattering processes.
 
The time-independent TDHFB equations can be easily obtained within the transformations: $\Phi_B(x,t)=\Phi_B (x)\exp (-i\mu_B t/\hbar)$, $\tilde{m} (x,t)=\tilde{m} (x)\exp (-i\mu_{\tilde{m}} t/\hbar)$ and $\Phi_I(x,t)=\Phi_I(x) \exp (-i \mu_I t/\hbar)$, where $\mu_B$, $\mu_{\tilde{m}}$, and $\mu_I$ are, respectively, the chemical potential of the condensate and of the anomalous density and of the impurity. Strictly speaking $\mu_{\tilde{m}}$ is also associated with the
thermal cloud density since $\tilde{n}$ and $\tilde{m}$ are related to each other by Eq. (\ref {eq6}). Then the static TDHFB equations read
\begin{widetext} 
\begin{subequations}\label{TC:td}
\begin{align}
&\mu_B {\Phi}_B= \left [-{\frac{\displaystyle\hbar^2}{\displaystyle 2m_B}}\Delta +\frac{1}{2} m_B \omega_{B x} ^2 x^2+g_B \left(\beta |\Phi_B|^2+2\tilde{n}+\gamma|\Phi_I|^2\right) \right]\Phi_B,  \label{TC:td1}
\\ 
&\mu_{\tilde{m}}\tilde{m} = \left[ -{\frac{\displaystyle\hbar^2}{\displaystyle 2m_B}}\Delta + \frac{1}{2} m_B \omega_{B x} ^2 x^2+g_B \left(2G \tilde {m}+2n+\gamma|\Phi_I|^2\right)\right]\tilde{m}, \label{TC:td2}
\\
&\bar\mu_I\Phi_I = \left[-{\frac{\displaystyle\hbar^2}{\displaystyle 2m_I}}\Delta + \frac{1}{2} m_I \omega_{I x} ^2 x^2+g_{B} (\gamma|\Phi_B|^2+\gamma\tilde{n})\right]\Phi_I. \label{TC:td3}
\end{align}
\end{subequations} 
\end{widetext} 
%Equations (\ref {TC:td}) have a trivial solution with $\gamma$=0.

To gain insight into the behavior of the thermal cloud and the anomalous densities in the BEC-impurity system at finite temperature, we solve numerically Eqs. (\ref{TC:td}) 
using the finite differences method.\\  
In the numerical investigation,  we use $a_0=\sqrt{\hbar/ m_B\omega_{B x}}$ and $\hbar \omega_{B x}$ as the length (the ground state extent of a single BEC-boson particle) 
and the energy units, respectively, and we end up with $\alpha=m_B/m_I$ being the ratio mass and $\Omega=\omega_{B x}/\omega_{I x}$. \\
The parameters are set to:  $N_I=1$ of ${}^{85}$Rb impurity atom, $N$=$10^5$ of ${}^{23}$Na bosonic atoms, $a_B$=3.4nm, $a_{IB}$=16.7nm, the transverse trapping frequency is $\omega_{B \perp}=2\pi \times500$ Hz\cite {Tim1}, the longitudinal trapping frequency is $\omega_{B x}=2\pi \times5$ Hz, $\gamma=4.91$ and $\Omega=0.2$.\\
%In practice, we can also speak about a self-localized state even if an impurity is focused in the center only by means of the interactions, $\Omega=0$.\\
\begin{figure}  [htbp]
\centerline{
\includegraphics[width=4.5cm,height=5cm,clip]{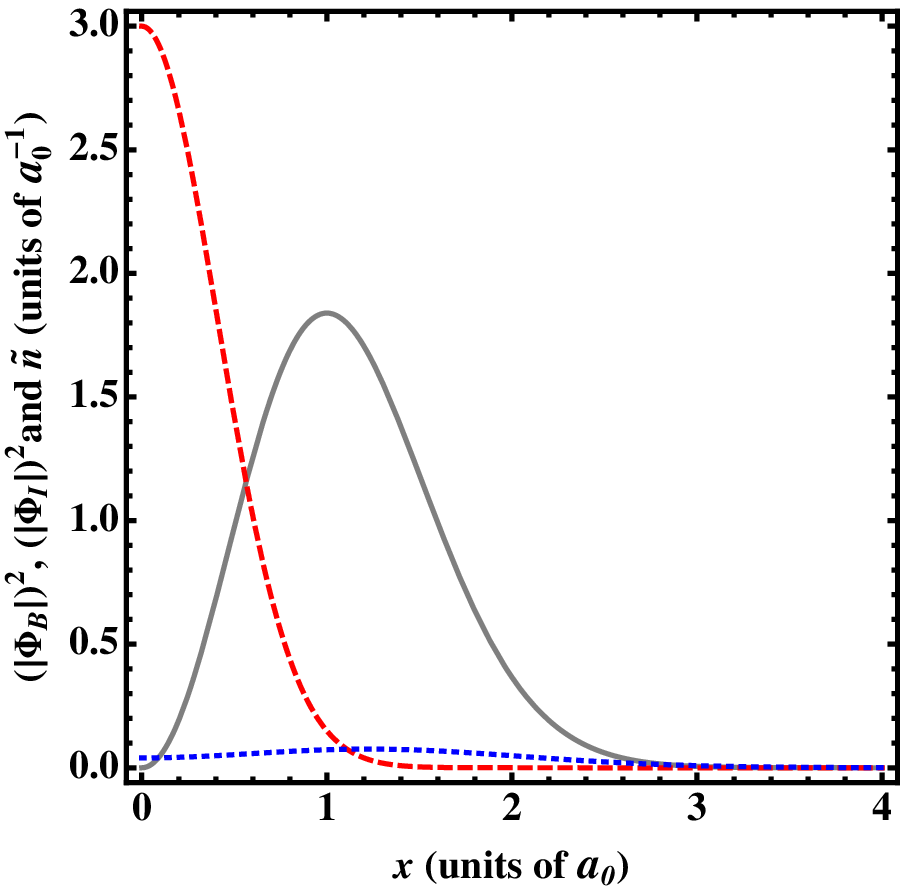}
\includegraphics[width=4.cm,height=5.cm,clip]{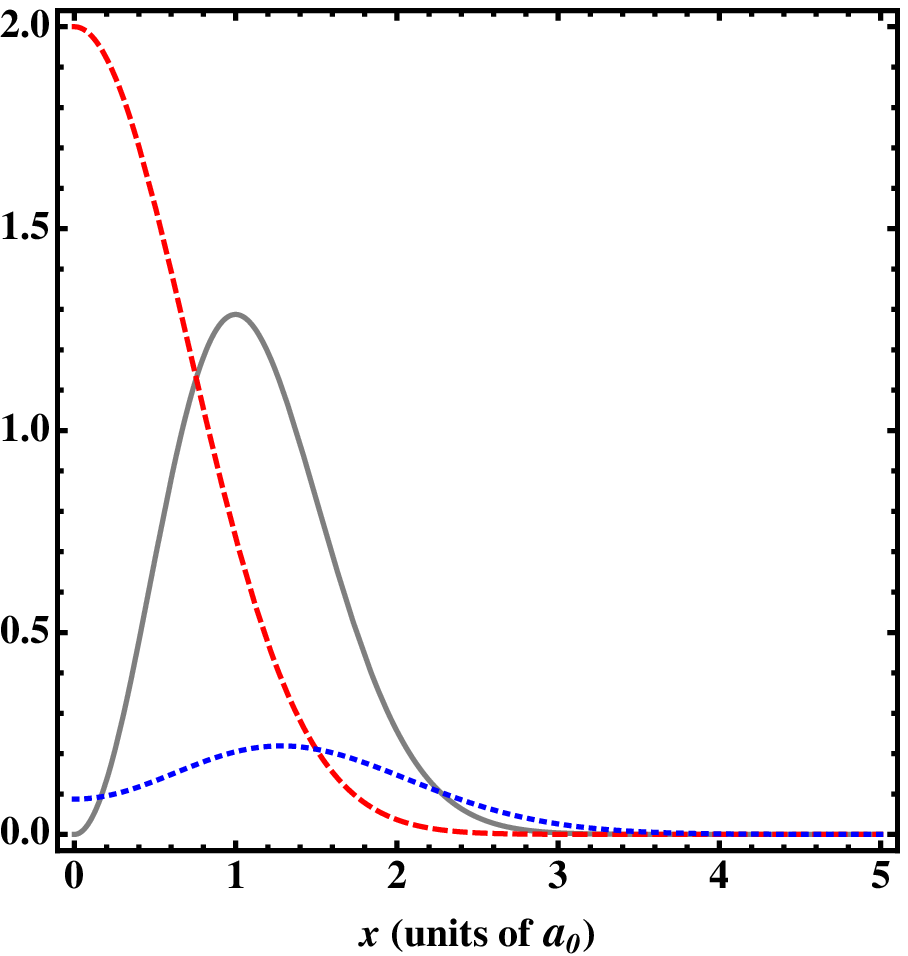}}
 \caption{(Color online) Condensed (gray lines), noncondensed ( red-dashed lines) and impurity (blue-dotted lines) densities as function of the radial distance for $\beta=0$ (left panel) 
  and for $\beta=1.1$  (right panel) for the above parameters.}
\label{profile} 
\end{figure}
Our numerical simulations show that for repulsive interactions, the condensate is distorted by the impurity and forms a dip near the center of the trap.
The impurity is focused inside the condensate forming a self-localized state as is illustrated in the left panel of Fig.\ref {profile} 
which is in good agreement with existing theoretical results.
One can see also from Fig. \ref {profile} (right panel) that the density of the condensate and of the impurity is lowered for $\beta=1.1$. 
In addition, the thermal cloud is deformed away from the impurity, as happens to the condensate cloud. 
The density of the impurity reduces and becomes less localized when the temperature grows as shown in the same figure. 
Indeed, this decay arises from the fact that at nonzero temperatures the condensate coexists with both a noncondensed cloud and an anomalous density composed of thermally excited
quasiparticles. Therefore, interactions between condensed and noncondensed atoms on the one hand and interactions of the impurity with atoms of the surrounding condensate on the other hand lead to dissipation, so that the impurity loses energy and delocalizes.

\begin{figure} [htbp] 
\centerline{
\includegraphics[width=8cm,height=8cm,clip]{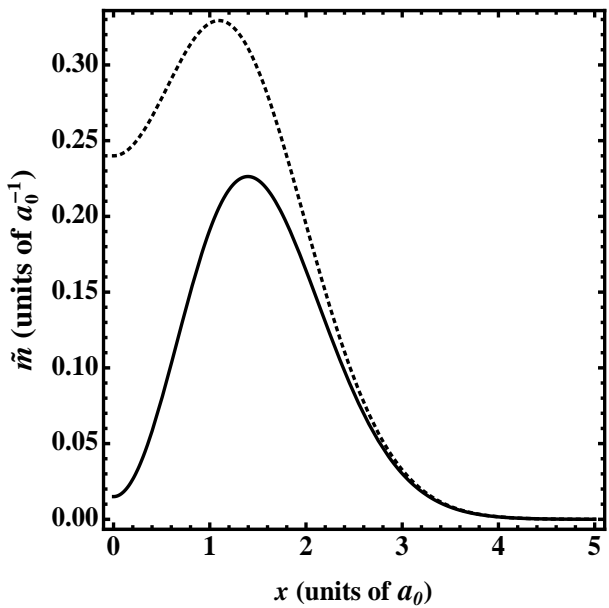}}
 \caption{Anomalous density as a function of the radial distance for $\beta=1.1$ 
   with the same parameters as in Fig. \ref {profile}. Solid lines: in the presence of the impurity. Dotted lines: without impurity.}
\label{profm} 
\end{figure}

A qualitative difference can be observed between anomalous density  with impurity and anomalous density without impurity. 
Figure.\ref{profm}, shows that the dip in the neighborhood of the center of the trap, 
which arises from the interactions between atoms of the condensate and those of the thermal cloud \cite{boudj2011}, becomes deeper in the presence of the impurity. 
This clearly confirms that the anomalous density is also distorted in an analogous manner with the condensate. 

%This vortex constitutes a new sort of vortices in BECs and known as the anomalous vortex \cite{boudj 2013}. 
%Note that this vortex reaches it maximum at intermediate temperature and disapears near the transition \cite{boudj 2013}.

\section{Effects of BEC fluctuations on the self-trapping impurity}\label{BECflu}

In order to study effects of BEC fluctuations on the self-trapping problem of weakly BEC-impurity interactions in the homogeneous case ($V_B=V_I=0$), 
it is convenient to linearize Eqs. (\ref{TC:td1}) and  (\ref{TC:td2}) by
considering the small deformations \cite{Tim4} $\delta\Phi_{B} =\Phi_{B}-1$ and $\delta\tilde{m}=\tilde{m}-1$ of the condensate and of the anomalous density, respectively. 
Assuming that $\delta\Phi_{B}$ and $\delta\tilde{m}$ are real for simplicity. The linear equations take the following forms
\begin{subequations}\label{Sl:td}
\begin{align}
&\left (-\frac{1}{2} \Delta + A\right)\delta\Phi_B  =-C|\Phi_I|^2,  \label{Sl:td1} \\ 
&\left (-\frac{1}{2} \Delta+ B\right)\delta\tilde{m}  =-C|\Phi_I|^2, \label{Sl:td2} \\
&\left [-\frac{1}{2} \Delta + \frac{\gamma}{\alpha} (2\delta\Phi_B+\delta\tilde{m})\right]\Phi_I  =\bar\varepsilon\Phi_I. \label{Sl:td3}
\end{align}
\end{subequations} 
where \\
$A=2+2(\beta-2)+\bar\mu_B$,\\
$B=2+4G+\bar\mu_{\tilde{m}}$,\\
$C=\gamma/\xi n$, \\
$\xi=\hbar/\sqrt {m_B ng_B}$ is the healing length,\\
$\bar\varepsilon=(\bar\mu_I-3\gamma/2)/\alpha$,\\
$\bar\mu_B=\mu_B/ng_B$, $\bar\mu_{\tilde{m}}=\mu_{\tilde{m}}/ng_B$ and $\bar\mu_I=\mu_I/ng_B$. \\
Equation.(\ref {Sl:td3}) constitutes a natural extention of that used in the literature \cite {Jack1,Tim4} since it contains the condensate and its fluctuation. 
The solution of  this equation allows us to study not only the self localizing problem at finite temperature 
but also enables us to see how the condensate fluctuation enhances the thermodynamics of the impurity such as the chemical potential and the compressibility.\\
It can be seen from Eqs. (\ref {Sl:td1}) and (\ref {Sl:td2}) that the linearization of Eqs. (\ref {TC:td}) is valid in the regime $C\ll1$.
The solution of Eqs. (\ref {Sl:td1}) and (\ref {Sl:td2}) is given in terms of the Green’s function $G(x)$. Inserting this solution into Eq.(\ref {Sl:td3})
with the assumption that $\delta\tilde{m}/\delta\Phi_B\ll1$ at low temperature,  one finds that $\Phi_I$ obeys the non-local non-linear Schr\"o\-dinger equation
\begin{equation}  \label{eq13}
\left [-\frac{1}{2} \Delta -2 \zeta\int dz' G(z,z')|\Phi_I(z')|^2\right]\Phi_I  =\bar\varepsilon\Phi_I,
\end{equation}
where $\zeta=\gamma C/\alpha$ is the self-trapping parameter.\\
Multiplying Eq. (\ref{eq13}) by $\Phi_I ^{*}(z)$, integrating over $z$, and making use of the normalization condition, we obtain
\begin{align} \label{eq14}
&\bar\varepsilon =\bar\varepsilon_{kin}+\bar\varepsilon_{def}\\
&=-\frac{1}{2} \int dz \Phi_I ^{*}(z)\Delta \Phi_I(z) \nonumber \\ 
&-\zeta \int dz \int dz' |\Phi_I(z)|^2 G(z,z')|\Phi_I(z')|^2, \nonumber
\end{align}
where $\bar\varepsilon_{def}$ is the energy gained by deforming the BEC.\\
To estimate the critical parameters for which self-trapping occurs, we insert the normalized Gaussian wavefunction 
$\Phi_I (z)=(1/\sqrt{\pi q^2})^{1/4}\exp (-z/2q)^2$. A straightforward calculation yields
 \begin{equation}  \label{enrgimp}
\bar\varepsilon=\frac{1}{4q^2}-\zeta f(q),
\end{equation}
where $f(q)= (1/2) \exp({-2q^2})\text{erfc}(\sqrt{2}q) $ with $\text{erfc (x)}$ being the complementary error function.\\
Equation (\ref {enrgimp}) provides a useful expression for the impurity chemical potential 
 \begin{equation}  \label{chimimp}
\bar\mu_I= \frac{\alpha}{4q^2}+\gamma \left[\frac{3}{2}-C f(q)\right].
\end{equation}
It is clearly seen from Eq. (\ref {chimimp})  that, for $q>1$, $\bar\mu_I$ is linearly increasing with $\gamma$.\\
Importantly, Eq.(\ref {chimimp}) shows that the variational impurity chemical potential differs by a factor of $3/2$ compared to the ordinary zero temperature case i.e without fluctuations.
We then infer that the presence of thermal fluctuations of the condensate leads to corrections of the chemical potential of the impurity.

The above chemical potential implies the following expression for the impurity compressibility $\kappa_I^{-1}=n^2 \partial \bar\mu_I/\partial n$:
 \begin{equation}  \label{comp}
\kappa_I^{-1}=\frac{1}{2}\frac{\gamma^2}{\xi} f(q).
\end{equation}
The compressibility (\ref {comp}) remains finite and increases with $\gamma$.

If $q\gg 1$, we can Taylor-expand $f$ as $f\approx1/2-\sqrt{2/\pi} q$. In this limit, the impurity energy $\bar\varepsilon=1/(4q^2)+\zeta q/\sqrt{2\pi}-\zeta/2$,
attains a minimum at $q=0.85\,\zeta^{-1/3}$. Therefore,  the self-trapping occurs for small $\zeta$ in quasi-1D BEC-impurity 
which is in agreement with the theoretical results of Ref. \cite {Jack1}. \\
We conclude that the condensate fluctuations do not have considerable effects on the occurrence of the self-tapping at low temperature.

It is worth noting that our model is also applicable in harmonically trapped BEC. 

\section{Solitons in the BEC-impurity system} \label{sol}
Our aim in this section is in a sense twofold. On the one hand, we aim to study the formation of matter-wave solitons in
BEC-impurity mixtures at finite temperature in an experimentally relevant and realizable setting. 
On the other hand, we are aiming to see what are the effects of the temperature or the dissipation on the generation of solitons.

What is advantageous in our model (\ref{TC:td}) is that the anomalous density is 
treated dynamically on the same footing as the condensate, which leads us to predict a new kind of soliton, namely, an “anomalous soliton”. 
This soliton occurs generically in the thermal equilibrium state of a weakly interacting Bose gas irrespective of the presence or not of the impurity.
At this point, one should mention that the previous analysis of parameter $\beta$ highlights the emergence of, at least, two different cases 
for BEC with repulsive interactions ( $g_B>0$): bright anomalous soliton for $0<\beta<1$ and dark anomalous soliton for $\beta >1$. 

To investigate in more detail the formation of solitons in a weakly repulsive BEC-impurity under the presence of thermal fluctuations, we consider a quasi-1D (elongated along the $x$ direction) geometry which is the most favorable for the appearance of solitons. 
Again, we solve numerically Eqs. (\ref{TC:td}), employing appropriate boundary conditions with the same experimental values corresponding to Fig.\ref {profile}.

\begin{figure}  [htbp]
\centerline{
\includegraphics[width=3.cm,height=5cm,clip]{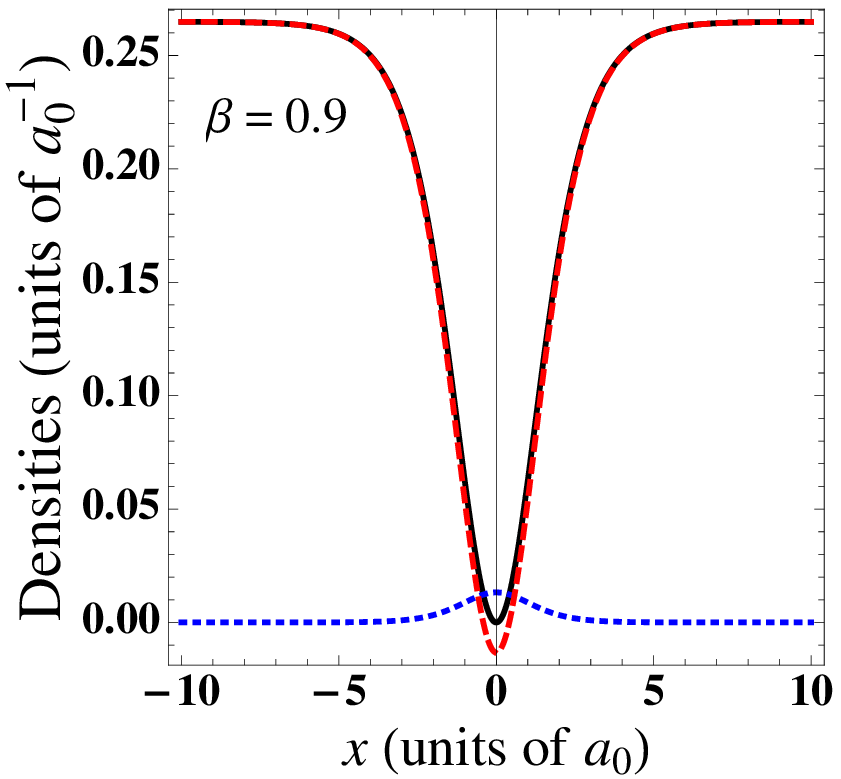}
\includegraphics[width=2.5cm,height=5cm,clip]{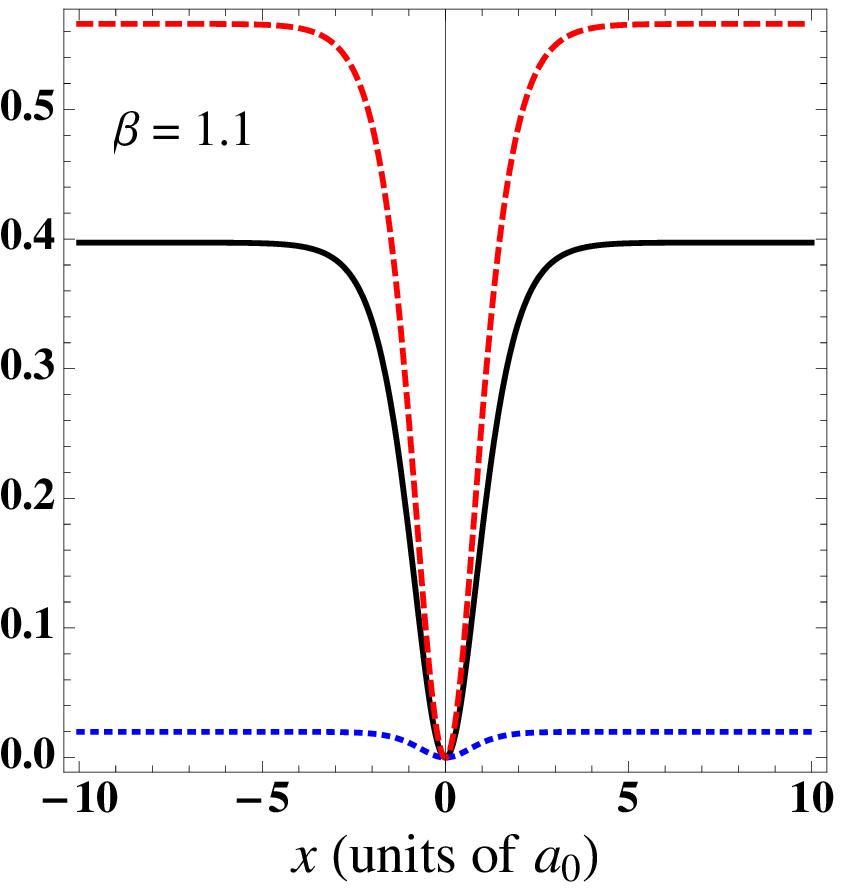}
\includegraphics[width=2.5cm,height=5.cm,clip]{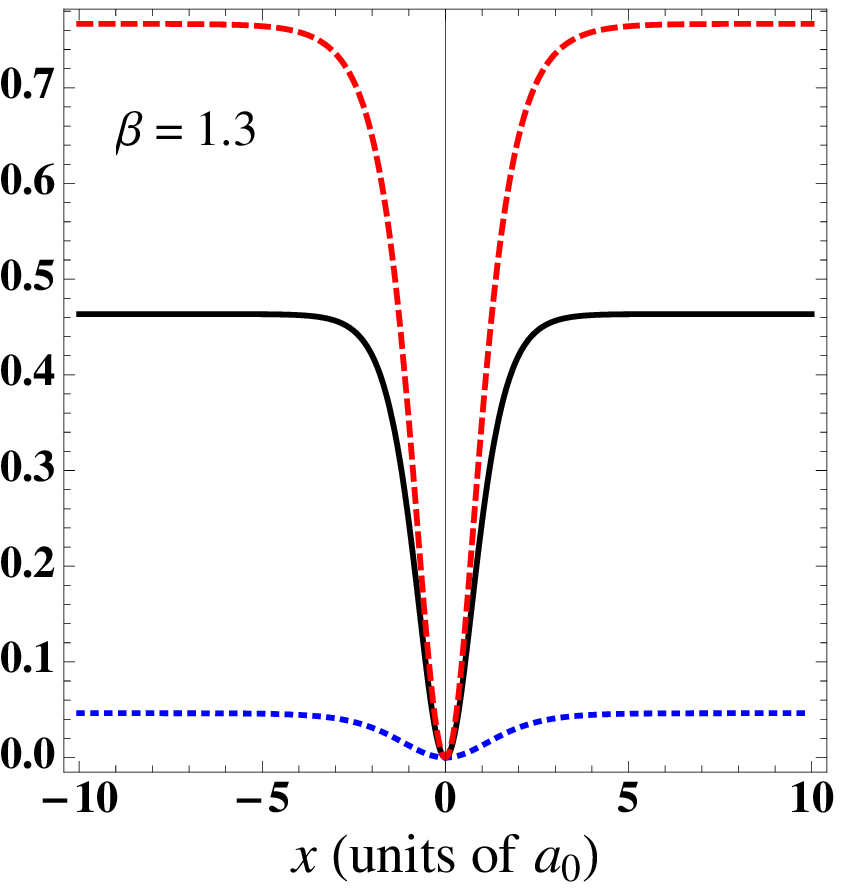}}
 \caption{(Color online) Density profiles for solitons in the BEC-impurity mixture with the same parameters as in Fig. \ref {profile}. Solid lines: ordinary soliton, red-dashed lines: impurity soliton.  Blue-dotted lines: anomalous soliton.}
\label{soli} 
\end{figure}

Figure.(\ref{soli}) depicts clearly  the formation of dark solitons in the condensed and the impurity components 
and a bright soliton in the anomalous density for $\beta=0.9$.
The situation is inverted for $\beta=1.1$ where a spontaneous dark anomalous soliton is generated, without any external forcing or perturbations
which is in good accordance with our previous analysis. This soliton becoming widespread and deep as temperature rises  
unlike to the condensed (ordinary) and impurity solitons where they become narrower and deeper at higher temperatures because they lose energy due to the dissipation.
A similar behavior has been predicted in Ref.\cite{karp1} for thermal solitons in a quasi-1D Bose gas. 
Also, a careful observation of the same figure shows that the impurity soliton is more deeper than the condensed one and the depth of these three solitons increases with increasing temperature.
Another important remark is that the impurity soliton is localized inside the ordinary one especially for values of $\beta >1$ and both solitons are localized in the core of the anomalous soliton.
Consequently, the width of the anomalous soliton is larger than that of the ordinary soliton whatever the range of the temperature. This is in fact natural since the anomalous soliton is related to
the thermal cloud and this latter surrounds the condensate as it was shown in earlier BEC experiments.

It is understood also that for the BEC-impurity with attractive interactions ($g_B<0$), bright anomalous solitons can be produced at higher temperature ( for $\beta >1$). 

%We turn now to analyze the dynamics of the anomalous solitons. Usually, solitons can be moved by shaking slightly the trapping potential.
%In such a situation, the anomalous soliton obeys the time-dependent TDHFB equations. 
%For $\beta=0.9$, we observe from Fig.\ref{As} that a bright anomalous soliton propagates with almost constant
%amplitude reflecting the robustness and the stability of this soliton during its evolution. This behavior persists also for dark anomalous solitons.

%\begin{figure}  [htbp]
%\centering
 %\includegraphics[scale=.9, angle=0]{AMS.eps}
 %\caption{(Color online) Evolution of the anomalous soliton for $\beta=0.9$ with the same parameters as in Fig.\ref {soli}. 
%Here we measure time in units of $\omega_{B x}^{-1}$, the position in units of $a_0$ and the density of $a_0^{-1}$}
%  \label{As}
%\end{figure}

An interesting question that begs to be asked is what kind of solitons will exist in the BEC-impurity mixture 
with attractive boson-boson interactions and repulsive boson-impurity interactions or inversely?
For example for Bose-Fermi gas mixtures, it has been shown that bright solitons are produced as a result of a competition between two interparticle
interactions: boson-boson repulsion versus boson-fermion attraction \cite{karp}. \\
The response to this question and others related to the formation and the behavior of soliton molecules in BEC-impurity systems will be given elsewhere.

\section{Conclusion} \label{conc}
In this paper we have derived from the time-dependent BV variational principle a set of coupled equations 
for the BEC-impurity mixture. These equations govern in a self-consistent way the dynamics of the condensate, the thermal cloud, 
the anomalous average and the impurity. 
The numerical simulations of the TDHFB equations in the harmonically trapped quasi-1D model showed that the thermal cloud and the anomalous density 
are distorted by the impurity as happens with the condensate. 
Additionally,  the impurity is reduced and becomes less localized with increasing temperature.

Furthermore, we have investigated effects of BEC fluctuations on the self-trapping impurity in homogeneous weak interaction regimes at low temperature. 
We have found that these fluctuations may enhance the chemical potential and the compressibility of the impurity, while they do not affect the occurrence
of the self-trapping state. We have shown that the self-trapping takes place for small values of $\zeta$ in agreement with the case of zero temperature.

Moreover, we have studied the formation of matter-wave solitons in repulsively quasi-1D BEC-impurity mixtures in the presence of thermal effects.
Our formalism reveals the formation of stable solitons. Depending of  parameter $\beta$, the system contains much more than the standard picture. 
A dark soliton is created in condensed and impurity parts of the system whereas a bright soliton is formed in the anomalous density.
A  dark anomalous soliton is willingly generated at higher temperatures without the need of any external perturbations or squeezing of the geometry. 
This anomalous soliton is shown to be stable and robust during its time evolution.
Our formalism allows us to explain the temperature dependence of the appearance of deep solitons in the BEC-impurity.

An important step for future theoretical studies in the finite-temperature regime is to fully include the interaction part of the impurity atoms in the total Hamiltonian of the system \cite{Legt}.
This permits us to study in a self-consistent way, within the TDHFB formalism, fluctuations of the impurity and their effects on the formation of solitons and vortices in such a system.
 
\section{Acknowledgments}
We gratefully acknowledge Tomi Johnson for  discussion and comments on this paper.
This work was supported by the Algerian government under Research Grant No. CNEPRU-D00720130045.

\end{document}